\documentclass[12pt]{article}
\usepackage{cite,epsfig,verbatim,rotating,subfigure}
\usepackage{times}
\usepackage{fullpage,setspace}
\usepackage{url}
\usepackage{algorithmic}
\usepackage{algorithm}
\usepackage[toc,page]{appendix}
\usepackage{wrapfig}
\begin{document}

\title{A Domain Decomposition Strategy for Alignment of Multiple Biological Sequences on Multiprocessor Platforms}

\author{Fahad Saeed\\
{\small \url{fsaeed2@uic.edu}}\\
\and
Ashfaq Khokhar\\
{ \small \url{ashfaq@uic.edu}}\\
\and
{\small\em Department of Electrical and Computer Engineering}\\
{\small\em University of Illinois at Chicago}\\
{\small\em Chicago IL, USA} }

\maketitle

\begin{abstract}
\thispagestyle{empty} \singlespacing

\textit{Multiple Sequences Alignment (MSA) of biological sequences
is a fundamental problem in computational biology due to its
critical significance in wide ranging applications including
haplotype reconstruction, sequence homology, phylogenetic
analysis, and prediction of evolutionary origins. The MSA problem
is considered NP-hard and known heuristics for the problem do not
scale well with increasing number of sequences. On the other hand,
with the advent of new breed of fast sequencing techniques it is
now possible to generate thousands of sequences very quickly. For
rapid sequence analysis, it is therefore desirable to develop fast
MSA algorithms that scale well with the increase in the dataset
size. In this paper, we present a novel domain decomposition based
technique to solve the MSA problem on multiprocessing platforms.
The domain decomposition based technique, in addition to yielding
better quality, gives enormous advantage in terms of execution
time and memory requirements. The proposed strategy allows to
decrease the time complexity of any known heuristic of $O(N)^x$
complexity by a factor of $O(1/p)^x$, where $N$ is the number of
sequences, $x$ depends on the underlying heuristic approach, and
$p$ is the number of processing nodes. In particular, we propose a
highly scalable algorithm, Sample-Align-D, for aligning biological
sequences using Muscle system as the underlying heuristic. The
proposed algorithm has been implemented on a cluster of
workstations using MPI library. Experimental results for different
problem sizes are analyzed in terms of quality of alignment,
execution time and speed-up.}

\end{abstract}

\newpage
\doublespacing
\section{Introduction}
\label{sec:intro}

Multiple Sequences Alignment (MSA) in computational biology
provides vital information related to the evolutionary
relationships, identifies conserved motifs, and improves secondary
and tertiary structure prediction for RNA and proteins. In theory,
alignment of multiple sequences can be achieved using pair-wise
alignment, each pair getting alignment score and then maximizing
the sum of all the pair-wise alignment scores. Optimizing this
score, however, is NP-complete~\cite{NP-hard} and dynamic
programming based solutions have complexity of $O(L^N)$, where $N$
is the number of sequences and $L$ is the average length of a
sequence. Such accurate optimizations are not practical for even
small number of sequences, thus making heuristic algorithms a
feasible option. The literature on these heuristics is vast and
includes widely used works, including Notredame et al.
\cite{tcoffee}, Edgar \cite{Muscle}, Thompson et al.
\cite{clustalw}, Do et al. \cite{probcons}, Lassmann et al.
\cite{kalign}, Sze et al. \cite{psalign}, Schwartz et al.
\cite{AMAP} and Morgenstern et al. \cite{dialign}. These
heuristics are complex combination of ad-hoc procedures with some
flavor of dynamic programming. Despite the usefulness of these
widely used heuristics, they scale very poorly with increasing
number of sequences.

The high computational costs and poor scalability of existing MSA
algorithms make the design of multiprocessor solutions highly
desirable. Also, recent advances in the sequencing techniques such
as pyrosequencing~\cite{pyro} are enabling fast generation of
large amount of sequence data. For example, at the time of writing
this paper, UniProtKB/Swiss-Prot contains 366226 sequence entries,
comprising 132054191 amino acids representing 11342 species.
Comparing this with less than 50k sequences in 1995, gives a
glimpse of exponential growth in biological data. If useful
research has to proceed, the performance of multiple alignment
systems has to scale up accordingly with the enormous amount of
data being generated.

The main goal of the work presented in this paper is to
investigate  domain decomposition strategies for biological data
computations. For the multiple sequence alignment problem
discussed in this paper, we use k-mer rank, a metric that depicts
similarity of a sequence compared to another sequence, to
partition the input data set into load balanced subsets.
Subsequently, we show how these decomposed subsets of sequences
can be aligned on multiple processors in a distributed fashion,
and glued together to get a highly accurate alignment of multiple
sequences. Our approach is capable of aligning a large number of
sequences (of the order of 20000 sequences \cite{SaeedKhokhar}),
with time complexity scaled down by a factor of $O(1/p)^4$,where
$p$ is the number of processors, achieving super-linear speedups
on multiprocessors without compromising quality of the alignment.

The rest of the paper is organized as follows. We start with a
brief problem statement and background information relevant to our
discussions in Section 2. Also, we discuss existing parallel
approaches to the MSA problem and identify their limitations. In
Section 3, we discuss the proposed domain decomposition based MSA
algorithm for aligning protein sequences. This is followed by a
rigorous analysis of the computation and communication costs.
Section 4 presents the experimental results and analyzes these
results in terms of alignment quality, execution time, memory
usage, and speedup. Section 5 presents the conclusions and
outlines future research.

\section{Problem Statement and Background Information}

We first define the Multiple Sequences Alignment (MSA) problem in
simplest form, without indulging with the issues such as scoring
functions, which are beyond the scope of this work. Let $N$
sequences be presented as a set $S = \{S_1,S_2, S_3, \cdots,
S_N\}$ and let $S^{'} = \{ S^{'}_{1}, S^{'}_{2} , S^{'}_{3},
\cdots, S^{'}_{N} \}$ be the aligned sequence set, such that all
the sequences in $S^{'}$ are of equal length, have maximum
overlap, and the total alignment score is maximized according to
some scoring mechanism suitable for the application.

The method followed in most of the existing multiple alignment
systems is that a quick pair-wise alignment of the sequences is
performed, giving a similarity matrix. This similarity matrix is
then used to build a guide tree, which is then used to perform a
progressive profile-profile alignment. Note that profile-profile
alignments are used to re-align two or more existing alignments.
Profile-profile alignment is a useful method as it can be used to
gradually add new sequences to already aligned set of sequences,
also referred to as progressive alignment. It can also be used to
maintain one fixed high quality profile and keep on adding
sequences aligned to that fixed profile~\cite{clustalw}. These are
the basic steps that are followed by almost all distance based
multiple sequence alignment methods~\cite{Notredame}.
Fig.~\ref{fig-msa} shows a generic MSA scheme, which is also used
in Clustalw~\cite{clustalw}. The first stage is a pair-wise
comparison of the sequences under consideration. The second stage
corresponds to the construction of guiding tree, which is later on
used in stage three to perform final profile-profile alignments.
To improve the alignment score, various iterative methods have
been introduced in the later stages, as in the Muscle
System~\cite{Muscle}.


\subsection{Related Research}
There have been numerous attempts to parallelize existing
sequential multiple sequences alignment systems.
Clustalw~\cite{clustalw} is by far the most parallelized multiple
sequence alignment system. James et. al. in~\cite{clustalpc}
parallelized Clustalw for PC clusters and distributed shared
memory parallel machines. HT Clustal is a parallel solution for
heterogeneous multiple sequence alignment and MultiClustal is a
parallel version of an optimized Clustalw~\cite{clustalht} Zola et
al.~\cite{tcoffeep} provided the first parallel implementation of
T-Coffee based on MPI. Different modules of the Muscle system have
also been parallelized~\cite{musclep}. Other parallelization
efforts include parallel multiple sequence alignment with
phylogeny search by simulated annealing by Zola et
al.~\cite{Zola-phy}, Multithreading Clustalw for multiple sequence
alignment by Chaichoompu et al.~\cite{clustalmt} and Schmollinger
et al. parallel version of Dialign~\cite{dialignp}.

Although there seems to be a considerable amount of effort to
improve the running times for aligning large number of sequences
using parallel computing, it must be noted that all the existing
solutions have been aimed at parallelizing different modules of a
known sequential system. Therefore the parallelism achieved has
been limited to the usage of the function being parallelized. None
of the existing parallel alignment approaches has been able to
exploit the data parallelism, simply because of the lack of a
domain decomposition strategy. A few attempts~\cite{21}\cite{22}
have also been made to cut each sequence into pieces and compute a
piecewise alignment over all the sequences to achieve multiple
sequences alignment. In~\cite{22}, each sequence is 'broken' in
half, and halves are assigned to different processors. The
Smith-Waterman~\cite{21} algorithm is applied to these divided
sequences.  The sequences are aligned using dynamic programming
technique, and then combined using Combine and Extend
techniques~\cite{21}. The Combine and Extend methods follow
certain models defined to achieve alignment of the combination of
sequences. These methods pay little or no attention to the quality
of the results obtained.  The end results have considerable loss
of sensitivity. The constraints in these methods are solely
defined by the models used, thus limiting the scope of the methods
for wide variety of sequences.

Domain decomposition has been pursued for a large number of
application in numerous fields, including elliptic partial
differential equations, image processing, graphics simulations,
fluid dynamics, astronomical and atmospheric
calculations~\cite{vipin}. Most of these applications, take
advantage of parallel processing by decomposing the data domains,
and using data parallel techniques to achieve high performance.

In this paper, we investigate a data parallel approach to align
multiple protein sequences, consequently \emph{decreasing}
computational effort in terms of time and memory while
\emph{improving} or obtaining quality comparable to other multiple
alignment systems.

\section{Proposed Distributed MSA Algorithm: Sample-Align-D}


In this section, we present details of the domain decomposition
strategy and the alignment algorithm, referred to as
Sample-Align-D. We also analyze the computation and communication
complexities of the proposed algorithm.

The proposed domain decomposition strategy draws its motivation
from the Sample-Sort approach~\cite{sampling-1} that has been
introduced to sort a very large set of numbers on distributed
platforms. The sorting and MSA problems share a common
characteristic, i.e., any correct solution requires implicit
comparison of each pair of data items. In Sample-Sort, a small
sample ($\ll N$) representing the entire data set is chosen over
distributed partitions using some sampling technique such as
Regular Sampling~\cite{sampling-2}. Then each item can be
independently compared and ranked against this sample. If the
sample is a true representative of the underlying data set, it
eliminates the need for explicit comparison of every item with the
entire set. This way an $N$ size sorting problem is reduced to
solving $p$ independent sorting problems of size $N/p$. We use a
similar sampling approach to the domain decomposition of sequences
over all the processors. In the case of sorting of integer
numbers, numerical values of the numbers are compared to compute
the rank of each number. In the case of multiple sequence
alignment problem, we need to identify a unique feature of the
sequence that could be used to compute the rank of each sequence,
in terms of degree of similarity with other sequences in the set.
This rank information can then be used to partition the smaller
input subsets based on similarity and align smaller subsets of
similar sequences independently. We propose to use k-mer
distance~\cite{edgar-k} as a metric to determine the similarity of
a sequence with any other sequence. Intuitively, the k-mer
distance between any two sequences is based on the relative
frequency of repetitive substrings of size $k$ in the sequences.
Edgar in~\cite{edgar-k} showed that k-tuple similarities correlate
well with fractional identity, and the small values of $k$ between
4 and 6 work well for biological sequences. For the sake of
completeness, in the following, we provide the formal definition
of k-mer distance.
\\\
\textbf{k-mer Rank:} Let's assume that a biological sequence is
represented by a string $X$ of $n$ characters taken from an
alphabet $\AA$ that contains $c$ different characters
$(a_1,\cdots,a_c)$. For the words of length $k$ (hence named
k-mers), there are $\epsilon=c^k$ such different words. We
represent the set of k-mers in $X$ by vector $c^X=(c^X_1,
\cdots,c^X_{\epsilon})$. The distance between string $X$ and any arbitrary string $Y$, of length $m$,
is calculated using $c^X_i$ and $c^Y_i$, the count of k-mer
occurrences in $X$ and $Y$. Now let $C^{XY}_i = min(c^X_i,c^Y_j)$
denote the common k-mer count.

\begin{equation}
F(X,Y)= \sum_{i=1}^{\epsilon} \frac{C^{XY}_i} {[min(n,m)-k+1]}
\end{equation}
\begin{equation}
d^{F(X,Y)}= -log(\Delta + F(X,Y))
\end{equation}

\noindent where $F$ is the fraction of common k-mers between $X$
and $Y$, and $d^F$ transforms this into a distance. $\Delta$, is a
small constant added to prevent logarithm of zero. Based on k-mer
distance, we define \emph{k-mer rank} as follows:

\begin{equation}
R_i= \frac{1}{N} \sum_{j=1}^N d^{F(i,j)}
\end{equation}

\noindent

An intuitive outline of the proposed distributed multiple sequence
alignment solution, referred to as Sample-Align-D, is given in
Algorithm~\ref{algo1}.

\begin{algorithm}
\caption{Sample-Align-D Intuitive Description}\label{algo1}
\begin{algorithmic}
\REQUIRE $p$ processor for computation \REQUIRE $N$ sequences of
amino acids $S_1, S_2, \cdots, S_N$: \ENSURE Multiple alignment of
$N$ sequences

\begin{enumerate}
    \item In parallel, calculate the global k-mer rank for each sequence in each processor
    \item Redistribute the sequences using the k-mer ranks
    such that sequences with similar k-mer ranks are accumulated on the same processor.
    \item In parallel, align the sequences on each processor using any sequential multiple sequence alignment (MSA) system
    \item Calculate the global ancestor using local ancestors produced by the local alignments at each processor in the previous step.
    \item In parallel, fine tune the alignment on each processor using the global ancestor

\end{enumerate}
\end{algorithmic}

\end{algorithm}

\subsection{k-mer Rank based Decomposing Domain}

Our aim is to decompose the data set into subsets such that the
sequences within a subset are more similar to each other than the
sequences in other subsets. Ideally, this can be accomplished by
partitioning the phylogenetic tree in a load balanced fashion such
that the partitions also minimize communication across processors.
The partitioning of a problem graph can be performed by making a
virtual grid over the graph structure~\cite{bokhari-1}. This works
well only if the problem graph is uniform. In MSA, the phylogentic
trees are rarely uniform. Thus partitioning with naive techniques
will lead to non-uniform loads. However, most of the existing
partitioning techniques for non-uniform problem
graphs~\cite{bokhari-2,part1,part2,part3,part4} cannot be used in
this case due to progressive alignment dependencies in MSA, as
discussed extensively in~\cite{tcoffeep}. In the following, we
outline a novel domain decomposition strategy that is based on
k-mer rank similarities.

There are three main parameters that may contribute to the
computation load  in the MSA problem. These include:  the number
of sequences, the length of the sequences, and the similarity rank
(that we call k-mer rank as discussed in the previous sections).
However, as our analysis will reveal in the later sections, the
lengths of the sequences do not contribute much computationally.
Hence we can safely neglect the length of the sequences for
load-balanced partitioning, and consider the k-mer rank and the
number of sequences for partitioning and mapping. We will use a
novel sampling based strategy to compute \emph{global} k-mer
ranks.


\subsubsection{Globalised k-mer Rank}

For a highly divergent set of sequences, k-mer rank computed for
each sequence locally on each processor using only $N/p$ sequences
would be different from the k-mer rank computed using all the $N$
sequences. In order to address this problem, we sample $k$
sequences from each processor such that the k-mer ranks of these
$k$ samples represent the ranks of the corresponding set of $N/p$
sequences, yielding a total of $k \times p$ samples. Collectively,
it is safe to assume that these $k \times p$ samples represent the
entire set of $N$ sequences. The k-mer rank based ordering of
these $k \times p$ sequences yields a phylogenetic tree of the
samples, which in turn represent all the sequences. Each processor
re-computes the k-mer ranks of its sequences using this global
sample. Subsequently, redistribution based on this new k-mer rank
also ensures that sequences accumulated in each processor are
'similar' to each other.

\begin{figure}[htb]
\begin{center}
\includegraphics[scale=0.5,angle=000]{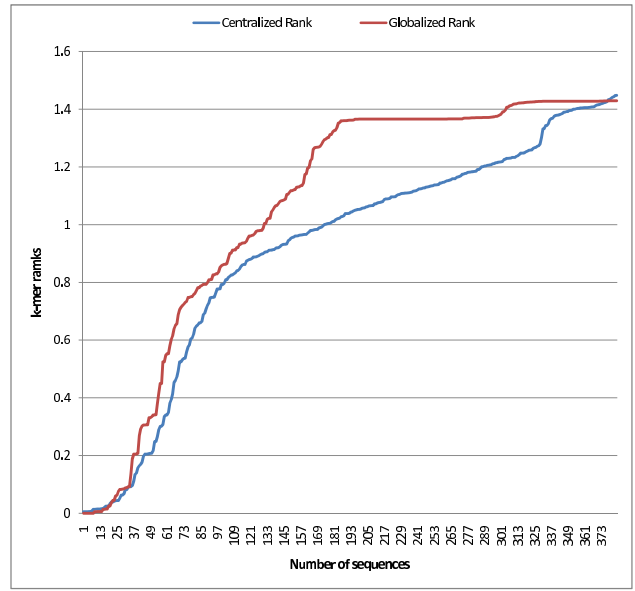}
 \caption{\small \label{fig-global} Globalized ranks
and Centralised Ranks } 
\end{center}
\end{figure}



In Fig.~\ref{fig-global}, we plot the k-mer ranks computed using
samples (referred to as globalized ranks) and using all the
sequences collectively (referred to as centralized ranks). As
depicted in this figure, the curves have high degree of
similarity. The statistics of the two approaches for 500 sequences
are presented in Table.~\ref{table-1}. As can be seen that the
standard deviation for the two sets of ranks is very low (~0.58).
This shows that the k-mer ranks for $N$ sequences computed using a
global sample is statistically indistinguishable from the k-mer
ranks computed using all the $N$ sequences.


\begin{table}
\begin{center}
\caption{\small \label{table-1} Comparison of the k-mer ranks
computed using global sample (globalized) and using the entire set
of sequences (centralized).}
\begin{tabular}{| l | c | r | }
\hline
   (Maximum, Minimum) Central & (1.44827, 0.0) \\ \hline
    Average Centralized & 0.722962 \\ \hline
   (Maximum, Minimum) Globalized & (1.46207,0.0) \\ \hline
    Average Globalized & 1.11302 \\ \hline
    Variance w.r.t. Centralized & 0.33190 \\ \hline
    Standard Dev. w.r.t Centralized & 0.576377  \\
    \hline
  \end{tabular}
\end{center}
\end{table}

\subsubsection{ Redistribution Based on Globalized k-mer Rank}
Each processor computes the k-mer ranks of its $w = N/p$ sequences
locally using all the $N/p$ sequences, and sorts the sequences
based on this local k-mer rank. Here $N$ is the number of
sequences in the input and $p$ is the number of processors. From
each of the $p$ locally sorted lists, $k = (p-1)$ evenly spaced
samples are chosen. The k-mer ranks of these $(p-1)$ samples
(pivots) divide the local set into $p$ ordered subsets.  The k-mer
ranks of these $p-1$ samples from each processor are gathered at
the root processor, yielding a set $Y$ of size $p(p-1)$ ranks.


This regular-sampled set $Y$ is sorted to compute the ordered list
$Y_1, Y_2, Y_3, \cdots , Y_{p(p-1)}$  determining the range of
k-mer ranks over all the processors. Then ranks $Y_{p/2},
Y_{{p+p/2}}, \cdots, Y_{(p-2)p+p/2}$ are chosen as pivots ($p$ in
total) dividing the k-mer rank range into $p$ buckets. These
pivots are then broadcast to all the processors. Each processor
sends the sequences having k-mer ranks in the range of bucket $i$
to processor $i$. 
For the bound on the size of the dataset in each processor after
redistribution, we refer to the analysis in Section 3.2.


\subsubsection{The Alignment}

Next, a sequential MSA program is executed on each processor.
Since our ultimate goal is to have a global alignment of all the
$N$ sequences, a procedure has to be devised to concatenate these
'chunks' of \emph{locally} (here \emph{locally} is defined as the
chunk of sequences that are aligned on a single processor) aligned
sequences so that the {\it global} alignment of multiple sequences
is achieved. Edgar in~\cite{Muscle2} has observed that multiple
sequences alignment for homologous sequences can be obtained by
aligning each sequence to the {\it root} profile. This approach is
similar to the one used in the PSI-BLAST, where a {\it known}
profile is used to align any query sequence with the sequences
that have generated the profile. This technique may also be
categorized as {\it template} based method, as observed by
Notredame in his recent work~\cite{Notredame}. We use a similar
concept along with domain decomposition of the sequences. We
extract the local ancestor from each processor after {\it locally}
aligning each subset in parallel. All of these local ancestors are
collected at the root processor and are aligned using a sequential
multiple sequence alignment algorithm. The ancestor of all the
local ancestors, referred to as the global ancestor, is then
broadcast to all the processors. Subsequently, the global ancestor
is used to perform a profile-profile alignment. That is, each set
of the locally aligned sequences (referred to as profile) in each
processor is aligned with the global ancestor profile.

In order to apply pair-wise alignment functions to profiles, a
Profile Sum of Pairs (PSP) scoring function must be defined. We
use the same PSP score as defined in~\cite{amino}
and~\cite{Muscle2}:


\begin{equation}
PSP^{xy}= \sum_{i}\sum_{j} f^{x}_{i} f^{y}_{j} \log(p_{ij}/p_i
p_j)
\end{equation}

Here $x$ and $y$ are the profiles being aligned, $i$ and $j$ are
the amino acid in profiles, $p_i$ is the background probability of
$i$, $p_{ij}$ is the joint probability of $i$ and $j$ aligned to
each other, $f^x_i$ is the observed frequency of $i$ in column $x$
of the first profile, and $x_G$ is the observed frequency of gaps
in that column. The same attributes are assumed for the profile
$y$. For our purposes, we will take advantage of PSP functions
based on the 200 PAM matrix~\cite{200PAM} and the 240 PAM VTML
matrix~\cite{VTML}. Some multiple alignment methods implement
different scoring functions such as Log expectation (LE)
functions, but for our purposes PSP scoring suffices. Of course,
future work on decomposition strategies might investigate such
functions in this context.

\begin{figure}[htb]
\begin{center}
\includegraphics[scale=0.4,angle=90]{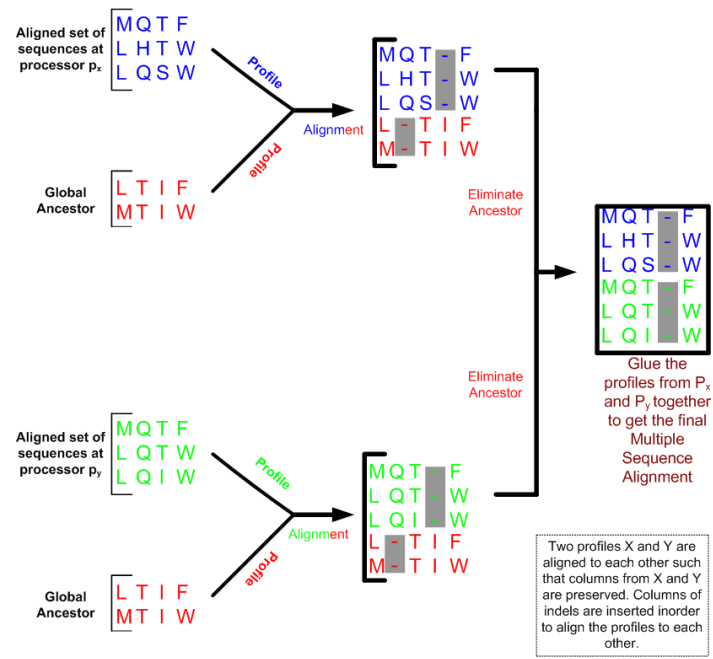}
\caption{\small \label{fig-profile2} Profile aligning with the
ancestor and combining sequence subsets.}
\end{center}
\end{figure}

This fine tuning step based on ancestor profile is depicted in
Fig.~\ref{fig-profile2}. For a highly divergent sequences set, we
propose an additional step, in which profiles can be added to the
root processor with respect to their similarity rank. This does
not change the computation or communication costs, but gives the
effect of 'profile-progressive' sort of gluing in the root
processor.

The summary of different steps in the Sample-Align-D Algorithm is
shown in Fig.~\ref{fig-summary},  and a detailed algorithmic
description is given in the Appendix as Algorithm~\ref{algo2}.

\begin{figure}[htb]
\begin{center}
\includegraphics[scale=0.5,angle=0]{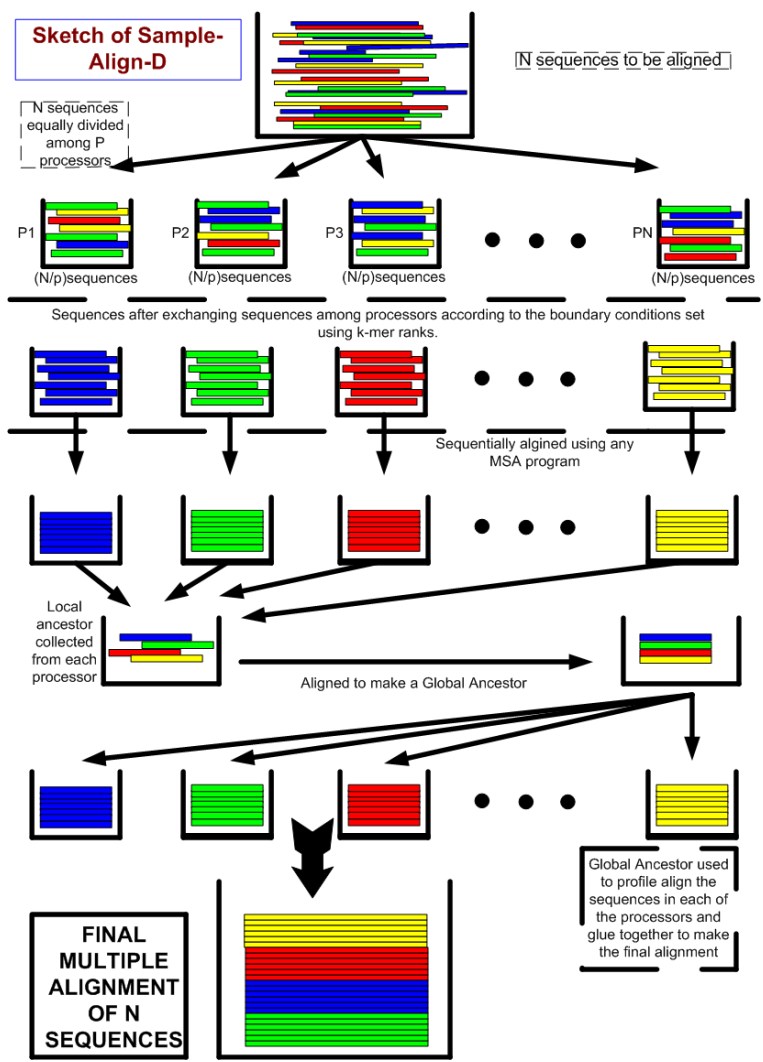}
\caption{\small \label{fig-summary} Summary of Sample-Align-D
procedure}
\end{center}
\end{figure}

\subsection{Analysis of Computation and Communication Costs}\label{sec-analysis}

For the computation and communication analysis we use a coarse
grained computing model such as $C^3$-model~\cite{Khokhar2} and
~\cite{Khokhar3}. Also, for analysis purposes, we assume that the
Muscle System~\cite{Muscle} is being used at each processor as the
underlying sequential multiple sequence alignment system. It must
be noted that the computation complexity of the alignment step
will vary depending on the sequential MSA system used for
alignment within each processor.

In the following analysis we assume that each processor has $w =
N/p$ sequences, where $N$ is the total number of sequences to be
aligned, and $p$ is the number of processors. The average length
of a sequence is $L$. In Table ~\ref{table-2}, we outline the
computation cost of each step of the algorithm and its memory
requirement.


\begin{table}

\begin{center}

\caption{\small \label{table-2} Computation Costs: }

\begin{tabular}{| l | c | c | }

\hline

STEP & O(Time) & O(Space)   \\ \hline

\scriptsize{k-mer rank computation on ($w=N/p$) sequences} & \scriptsize{ $w^{\mbox{2}} L $}& \scriptsize{ $ w+L $}\\

\scriptsize{Sorting of $N/p$ sequences based on k-mer rank} & \scriptsize{$w \log w $}& \scriptsize{ $ \log w $}\\

\scriptsize{Sample $ k = p-1 $ sequences }& \scriptsize{$w$} & \scriptsize{$p$ }\\

 \scriptsize{k-mer rank computation of  $(k \times p)$ sequences in root processor} & \scriptsize{$p^{\mbox{4}} L$} & \scriptsize{$p^{\mbox{2}} +L$}   \\

 \scriptsize{Sorting of $k \times p$ sample k-mer ranks} & \scriptsize{ $(k \times p) \log (k \times p)$} & \scriptsize{$ \log (k \times p)$}   \\

\scriptsize{k-mer rank computation of each of ($w=\frac{N}{p}$) sequences against $k \times p$ samples } & \scriptsize{$w[(k \times p+1)^{\mbox{2}}L$] }& \scriptsize{$w(k \times p+L)$}    \\

 \scriptsize{Muscle executed on ($w=\frac{N}{p}$) sequences in parallel }& \scriptsize{$w^{\mbox{4}}$ +$wL^{\mbox{2}}$       }  &\scriptsize{ $w^{\mbox{2}}$+$L^{\mbox{2}}$ }   \\

 \scriptsize{Ancestor extraction from each of the  $p$ processors + export to the root processor }& \scriptsize{$p^{\mbox{2}}$  }        & \scriptsize{$p^{\mbox{2}}$ }            \\

 \scriptsize{Muscle executed on local ancestors ($p$ elements)} & \scriptsize{ $(p)^{\mbox{4}}$+$(p)L^{\mbox{2}}$ }& \scriptsize{$(p)$ $^{\mbox{2}}$ + $L^{\mbox{2}}$ }  \\

 \scriptsize{Profile alignment with all combined aligned sequences on each of the processor} &\scriptsize{ $wL^{\mbox{2}}$ }& \scriptsize{ $ w$ }  \\ \hline

\small{TOTAL Computation Cost (for $w = \frac{N}{p}$)} & \small{$O((\frac{N}{p}) ^{\mbox{4}}$ + $(\frac{N}{p}) L^{\mbox{2}}$) } & \small{$O((\frac{N}{p})^{\mbox{2}}$ +$L^{\mbox{2}}$ ) }\\
    \hline

\end{tabular}
\end{center}
\end{table}

\subsection{Communication Cost}
The communication overhead is an important factor that dictates
the performance of a distributed message passing parallel system.
If the communication overhead is much higher than the computation
cost, the performance of the system is limited. Fortunately, the
communication cost of our system is much less than the cost of the
alignment. Essentially, the proposed Sample-Align-D algorithm has
two rounds of communication. In the first round, a small set of
samples is collected at the root processor and a set of pivots is
broadcast from the root processor. In the second round, sequences
are redistributed to achieve better alignments and balanced load
distribution. For the analysis of the communication costs we have
adopted the coarse grained computation model~\cite{Khokhar2}
and~\cite{vipin}. However, we ignore the message start up costs
and assume unit time to transmit each data byte.

We have assumed the Regular Sampling strategy~\cite{sampling-2}
because of its suitability to our problem domain. Some of the
reasons are :

\begin{enumerate}

\item The strategy is independent of the distribution of original
data, compared to some other strategies such as Huang and Chow
\cite{sampling-1}.

\item It helps in partitioning of data into ordered subsets of
approx. equal size.  This presents an efficient strategy for load
balancing as unequal number of sequences on different processors
would mean unequal computation load, leading to poor performance.
In the presence of data skew, regular sampling guarantees that no
processor computes more than $(2\frac{N}{p})$
sequences~\cite{sampling-2}.

\item It has been shown in~\cite{sampling-2} that regular sampling
yields optimal partitioning results as long as $N>p^3$, i.e., the
number of data items $N$ is much larger than the number of
processors $p$, which would be a normal case in the MSA
application.

\end{enumerate}

\subsubsection{ First Communication Round }

Assuming $k = p-1$, i.e., each processor chooses $p-1$ samples,
the complexity of the first phase is \textit{$O(p^{\mbox{2}}L$)+
$O(p \log p)+ O(k \times p \log p)$}, where $O(p^{\mbox{2}}L)$ is
the time to collect $p(p-1)$ samples  of average length $L$ at the root
processor, $O(p \log p)$ is the time required to broadcast $p-1$
pivots to all the processor and $(k \times p \log p)$ is the time
required to broadcast $k \times p$ sequences to all the
processors.

\subsubsection{Second Communication Round}

In the second round each processor sends the sequences having
k-mer rank in the range of bucket $i$ to processor $i$. Each
processor partitions its block into $p$ sub-blocks, one for each
processor, using pivots as bucket boundaries. Each processor then
sends the sub-blocks to the appropriate processor. The sizes of
these sub-blocks can vary from 0 to $\frac{N}{p}$ sequences
depending on the initial data distribution. Taking the average
case where the elements in the processor are distributed
uniformly, each sub-block will have $\frac{N}{p^2}$ sequences.
Thus this step would require $O(\frac{N}{p})$ time assuming an
all-to-all personalized broadcast communication
primitive~\cite{Khokhar3}. However, in the following we show that
based on regular sampling no processor will receive more than
$2\frac{N}{p}$ elements in total in the worst case. Therefore
still the overall communication cost will be $O(\frac{N}{pL})$.



Let's denote the pivots chosen in the first phase by the array:
$y_1, y_2, y_3,\cdots,y_{p-1}$. Consider any processors $i$, where
$1 < i < p$. All the sequences to be processed by processor $i$
must have k-mer rank $> y_{i-1}$ and $\leq y_i$. There are
$(i-2)p+ \frac{p}{2}$ sequences of the regular sample which are
$\leq y_{i-1}$, implying that there are at least $lb
=((i-2)p+\frac{p}{2})\frac{N}{p^2}$ sequences in the entire data
that have k-mer rank $\leq y_{i-1}$. On the other hand, there are
$(p-i)p-\frac{p}{2}$ sequences in the regular sample that have
k-mer rank $>y_i$. Thus, there are $ub =
((p-i)p-\frac{p}{2})\frac{N}{p^2}$ sequences of $N$ which are
$>y_i$. Since the total number of sequences is $N$, at most
$N-ub-lb$ sequences will get assigned to processor $i$. It is easy
to show that this expression is upper bounded by $2\frac{N}{p}$.
The cases for $i=1$ and $i=p$ are special because the pivot
interval for these two processors is $\frac{p}{2}$. The load for
these processors will always be less than $2\frac{N}{p}$. Due to
page limitations, we refer to~\cite{sampling-2} for further
details of the analysis.

The collection of $p$ local ancestors at the root processor and
broadcast of the global ancestor costs $O(L \log p)$ communication
overhead each. Therefore the communication cost is: $O(p^2L) + O(p
\log p) + O(\frac{N}{pL}) + O(L \log p)$.

The total asymptotic time complexity $T$ of the algorithm would
be:


\begin{equation}
\footnotesize{Computation Costs}= O(\frac{N}{p})^4 +
O(\frac{N}{p})L^2
\end{equation}
\begin{equation}
\footnotesize{Communication Costs}= O(p^2L)+ O(p \log p) +
O(\frac{N}{pL}) + O(L \log p)+ O(k \times p \log p)
\end{equation}
\begin{equation}
T \approx O((\frac{N}{p})^4 + (\frac{N}{p}) L^2) + (p^2L) +
(\frac{N}{pL})
\end{equation}

Next we briefly comment on the scalability of the Sample-Align-D
algorithm. We will use the
\emph{isoefficiency} metric~\cite{vipin} to show that
Sample-Align-D is highly scalable. For this we first define two important terms:
problem size defined as the number of basic computation steps to solve a problem on a
single processor using the best sequential algorithm; overhead function
is defined as the cost, that is not incurred by the fastest known
sequential algorithm. We denote
problem size with $W$, and overhead function with $T_o(W,p)$.

\begin{equation}
W=\frac{E}{E-1} \times T_o(W,p)
\end{equation}

where $E$ denotes the efficiency and let $K = \frac {E}{E-1}$. The
overhead function of Sample-Align-D:

\begin{equation}
T_o(W,p)\approx (\frac{N}{p})\times L + p^{2}\times L
\end{equation}


It is easy to show that asymptotically the iso-efficiency of
Sample-Align-D is $\Theta(p^2)$, i.e., the number
of sequences shall increase by a factor of $p^2$ to maintain the
efficiency with increasing number of processors.

\section{Performance Evaluation}

The performance evaluation process has been divided into two parts: the
first part deals with the quality assessment, and the second part
deals with traditional HPC metrics such as execution time,
scalability, memory requirements, etc. The performance evaluation
of the Sample-Align-D algorithm is carried out on a Beowulf
Cluster consisting of 16 Intel Xeon processors, each running at
2.40GHz, with 512KB cache and 1GB DRAM memory. As for the
interconnection network, the system uses Intel Gigabit network
interface cards on each cluster node. The operating system on each
node is Fedora Core 7(kernel level:2.6.18-1.2798.fc6xen).


\subsection{Quality Assessment}
The quality assessment in our case posed a considerable challenge
because most of the existing benchmarks such as
BaliBase~\cite{balibase} and Prefab~\cite{Muscle} used in the
literature are of very small sizes. Therefore they are not
effective in evaluating any {\it sampling} based approach or
domain decomposition based distributed approach. Also, other
parallel approaches to multiple alignment do not have any
decomposition strategy, making the quality of the parallel version
similar to the sequential version. Hence, a quality assessment
criterion for data parallel multiple alignment methods was not
available. In addition to assessing quality using these
traditional benchmarks, we have also formulated a method that can
be used to access the quality of the alignment produced by
distributed or data parallel MSA approaches.

\subsubsection{Quality Assessment using Traditional Benchmarks}
Traditional benchmarks such as BaliBase and Prefab are quite
comprehensive in terms of types of sequences contained in these
benchmarks. BaliBase, for example, has five basic categories and
covers most of the scenarios when making multiple sequence
alignments~\cite{tcoffee}. For the evaluation of multiple sequence
alignment programs, Balibase is divided into 5 hierarchical
reference sets:
\begin{itemize}
\item Ref1 for equi-distant sequences with various levels of
conservation, \item Ref2 for families aligned with a highly
divergent "orphan" sequence, \item Ref3 for subgroups with $<25\%$
residue identity between groups, \item Ref4 for sequences with
N/C-terminal extensions, and \item Ref5 for internal insertions.
\end{itemize}

Tables~\ref{table-3}, ~\ref{table-4} and ~\ref{table-5} compare
the quality of Sample-Align-D with different sequential algorithms
in terms of quality metrics, Q-Score, and TC-Score, used in
Balibase and Prefab benchmarks, respectively. The score for the
sequential algorithms have been derived from~\cite{Muscle}. The
Sample-Align-D was executed on a 4-processor system, therefore
corresponds to a 4 factor domain decomposition.


\begin{table}
\begin{center} \caption{\small \label{table-3} BaliBase Scores }
\begin{tabular}{| l || c || r | }
\hline
   Method      & Q & TC   \\ \hline \hline
   Muscle & 0.896&0.747 \\ \hline
   T-Coffee & 0.882&0.731 \\ \hline
   NWNSI(MAFFT)& 0.881& 0.722 \\ \hline
   Clustalw & 0.860&0.690 \\ \hline
   \textbf{Sample-Align-D}&\textbf{0.858}&\textbf{0.720}   \\ \hline
   FFTNSI(MAFFT) & 0.844&0.646  \\
    \hline
  \end{tabular}
\end{center}
\end{table}

\begin{table}
\begin{center}
\caption{\small \label{table-4} Prefab  Q-Scores }
\begin{tabular}{| l || c | }
   \hline
   Method      & Q-Score(All)    \\ \hline \hline
   Muscle & 0.645 \\ \hline
   \textbf{Sample-Align-D}&\textbf{0.623}   \\ \hline
   T-Coffee & 0.615 \\ \hline
   NWNSI(MAFFT)& 0.615 \\ \hline
   FFTNSI(MAFFT) & 0.591  \\ \hline
   Clustalw & 0.563 \\
   \hline
  \end{tabular}
\end{center}
\end{table}

In all the tests for quality assessment using benchmarks, it can
be seen that Sample-Align-D preformed very close to the Muscle
system. This is because Sample-Align-D was implemented with Muscle
System as the underlying sequential MSA algorithm at each
processor. Therefore, the quality obtained is limited by the
quality of the underlying alignment system.


\begin{table}
\begin{center}
\caption{\small \label{table-5} BaliBase Q-Scores on subsets}
\begin{tabular}{| l || c || c || c || c ||  r | }
\hline
   Method      & Ref1 & Ref2 & Ref3 & Ref4 & Ref 5   \\ \hline \hline
   \textbf{Sample-Align-D}&\textbf{0.882}&\textbf{0.932}&\textbf{0.800}&\textbf{0.872}&\textbf{0.804}   \\ \hline
   Muscle & 0.887& 0.935&0.823 & 0.876 &0.968 \\ \hline
   T-Coffee & 0.866& 0.934 & 0.787 & 0.917 & 0.957 \\ \hline
   NWNSI & 0.867 & 0.923 & 0.787 & 0.904 & 0.963 \\ \hline
    Clustalw & 0.861 &0.932 & 0.751 & 0.823 &0.859 \\ \hline
    FFTNSI & 0.838 &0.908 & 0.708 & 0.793 &0.947  \\
    \hline
  \end{tabular}
\end{center}
\end{table}

\subsubsection{Quality Assessment with Increasing Degree of Decomposition}

While the quality of alignment produced by the Sample-Align-D
algorithm for benchmark data sets is comparable to the other
systems, due to the small sizes of these benchmarks, we could not
evaluate the quality against several desired parameters that are
typical of a distributed environment such as sample size, number
of partitions, average length of sequences, etc. In the following,
we first outline the assessment procedure that allows us to
evaluate the quality while changing different parameters, and then
present performance results. In this subsection, we compare only
with the Muscle System.

\begin{figure}[htb]
\begin{center}
\includegraphics[scale=0.4,angle=0]{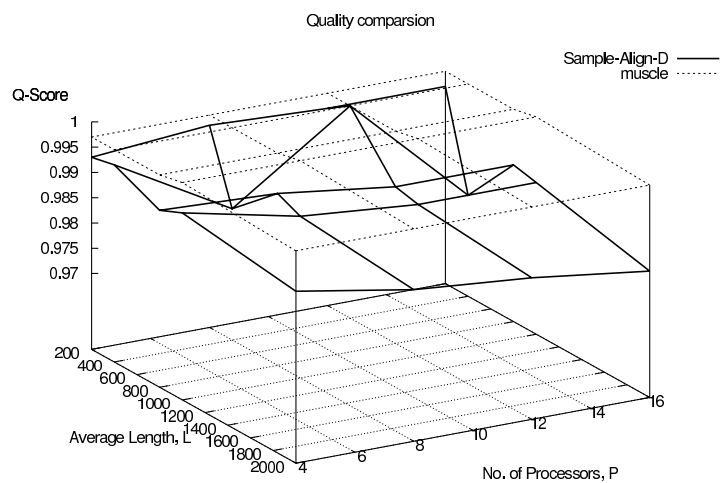}
\caption{\small \label{fig-result1} Quality comparison, with
varying average length of the sequences.}
\end{center}
\end{figure}

Using the Rose sequences generator~\cite{Rose} we have generated
23 sets of sequences, with their corresponding 'true' alignment,
while changing the following three parameters: the length of the
sequences, the number of sequences, and the phylogenetic distance
of the sequences. The length of the sequences in our tests varied
from 100 to 2000, the number of sequences varied from 100 to 20000
and the average phylogenetic distance varied from 100 to 1000.
These values are typical of biological sequences in existing
databases. Each of the set was aligned using Sample-Align-D with
different number of processors, and 'true' alignment obtained from
the Rose system was used as a benchmark. The metrics used for the
the assessment are Q-Score~\cite{Muscle},
TC-Score~\cite{balibase}, Modeler score~\cite{sauder}, Cline
(Shift) score~\cite{cline}, and Sum of Pairs (SP) score. The
quality scores obtained by the Sample-Align-D Algorithm are
compared with the Muscle System scores (only Q-scores are reported
in this paper).

The experiments were conducted with different data sets while
keeping two of the parameters held constants\footnote{Constants
held for the experiments are: Length =200, Number of sequence=200
and phylogenetic distance =100}. These parameters include number
of sequences, phylogenetic distance, and average sequence length.
The experiments were also conducted on different machine sizes to
study scalability and the effect of degree of domain decomposition
on quality of alignment.



\begin{figure}[htb]
\begin{center}
\includegraphics[scale=0.4,angle=0]{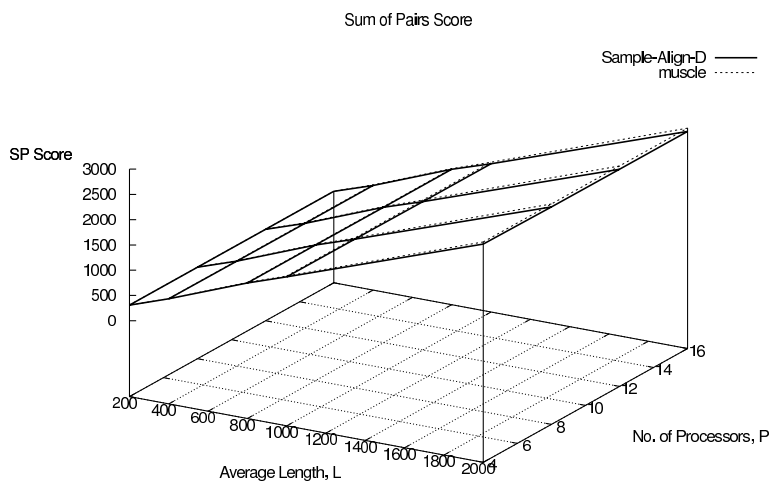}
\caption{\small \label{fig-result3} Sum of Pairs (SP) score, with
varying average length of the sequences.}
\end{center}
\end{figure}

Fig.~\ref{fig-result1} depicts performance in terms of Q-score
with increasing number of processors, while increasing average
length of sequences from 200 to 2000. It can be seen that the
increase in the number of processors to 16 didn't effect the
Q-scores. The Q-scores correlated very well with that of the
Muscle System. The scores remained above 0.97 for average length
of 2000. There was virtually no difference observed in the SP
scores computed for the Muscle system and Sample-Align-D for
respective pair of length and number of processors used, as shown
in Fig.~\ref{fig-result3}.

\begin{figure}[htb]
\begin{center}
\includegraphics[scale=0.4,angle=0]{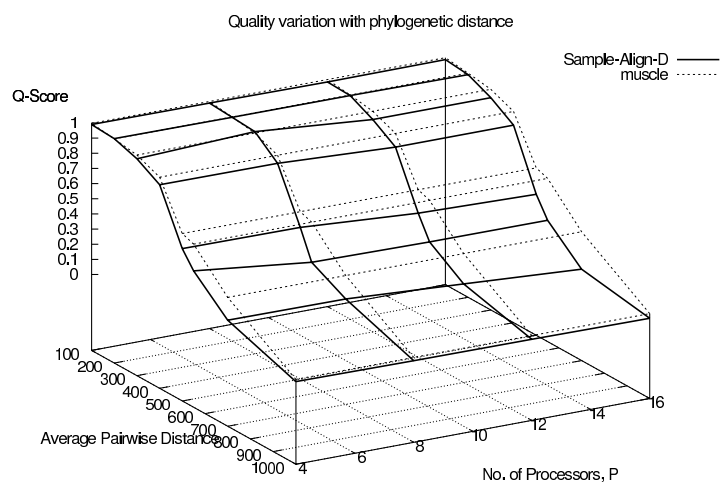}
\caption{\small \label{fig-result4} Q-score, with varying average
pairwise distance of sequences.}
\end{center}
\end{figure}

\begin{figure}[htb]
\begin{center}
\includegraphics[scale=0.4,angle=0]{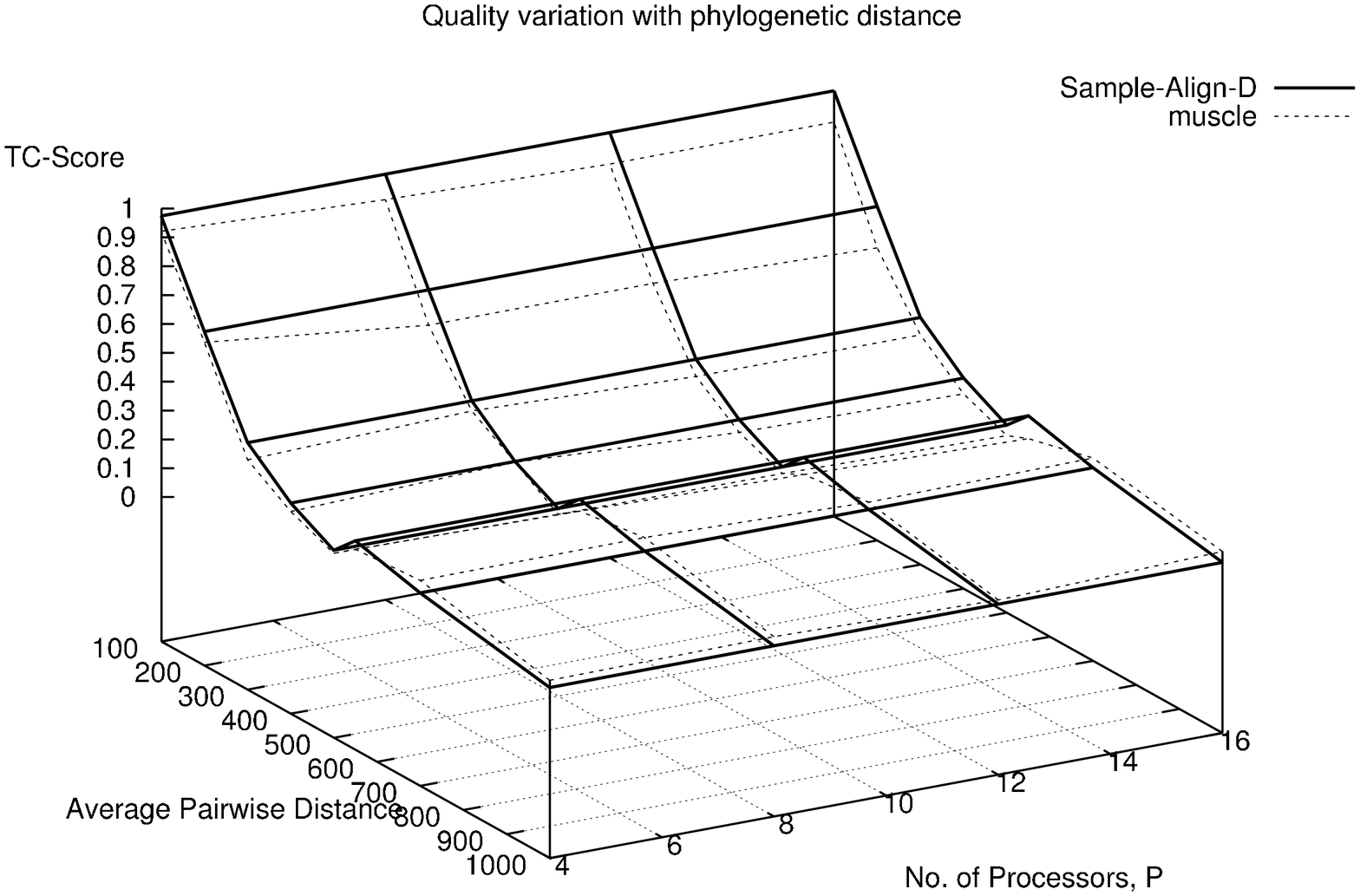}
\caption{\small \label{fig-result5} TC-score, with varying average
pairwise distance of sequences.}
\end{center}
\end{figure}

The quality with respect to the phylogentic distance is probably
the most important criterion. Theoretically, MSA systems should be
able to give good multiple alignments for increasing pairwise
sequence distance. After all, the relationship between distance
species would reveal the phylogenetics of the species. As pointed
out in~\cite{Lass}, all automatic multiple alignment systems
perform poorly with increasing pair wise distance between the
sequences. Different quality metrics calculated while increasing
the phylogentic distance from 100 to 1000 are shown in
Fig.~\ref{fig-result4} and Fig.~\ref{fig-result5} for Q-Scores and
TC-Scores, respectively. As can be seen from these figures, the
quality in fact decreased with increasing pairwise distance. Our
investigation for this criteria, however, was not to rectify the
quality issues with increasing pairwise distance, but to see the
correlation between the underlying MSA system and the effects of
the domain decomposition strategy.

\begin{figure}[htb]
\begin{center}
\includegraphics[scale=0.4,angle=0]{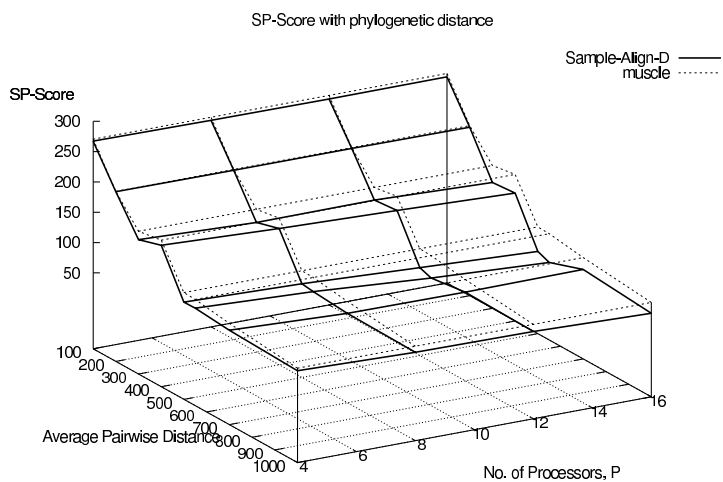}
\caption{\small \label{fig-result7} SP Score, with varying average
pairwise distance of sequences.}
\end{center}
\end{figure}

As can be seen from Fig.~\ref{fig-result4}, the Q-scores
correlated extremely well with the Q-scores of the Muscle System.
The TC-scores of the Sample-Align-D depicted in
Fig.~\ref{fig-result5} show slight \emph{increase} in the quality
when compared with TC scores of the Muscle System. The increase
can be attributed to the decomposition strategy which is more
inclined towards conserving columns in the multiple alignment,
owing to profile alignments. The decrease in the alignment scores
in general with increasing phylogenetic distance, is due to the
decrease of the alignment quality obtained from the underlying MSA
system. Without loss of generality, the quality of a decomposition
based MSA can be expected to correlate well with the underlying
sequential multiple alignment system that may perform well in
terms of quality. Fig.~\ref{fig-result7} shows the quality
performance in terms of SP score.

\begin{figure}[htb]
\begin{center}
\includegraphics[scale=0.4,angle=0]{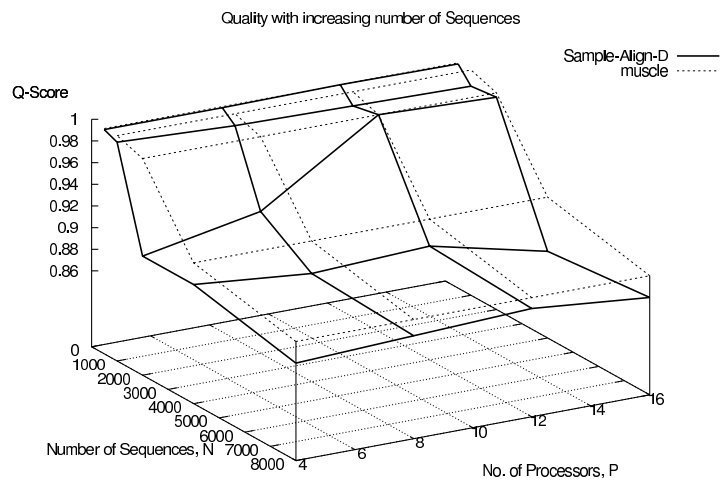}
\caption{\small \label{fig-result8} Quality of alignment, with
increasing number of sequences.}
\end{center}
\end{figure}

The assessment of the quality of alignment with increasing number
of sequences is also important. As before, we are interested in
the relative quality of alignment obtained after decomposition.
Fig.~\ref{fig-result8} shows the quality in terms of Q-Scores for
different sizes of the sequences set. The quality of the
Sample-Align-D Algorithm based alignments strongly correlates with
the quality of the alignments obtained by the Muscle system. It
must be noted that we are reporting quality for up to 8000
sequences, because the sequential Muscle System was unable to
process larger datasets.

\subsubsection {Example of Application to Serine/Threonine Kinases}

The purpose of a multiple sequence alignment system is to observe
and study the conservation of domain and motifs. To illustrate
this, we present here an example Fig.~\ref{fig-example} that
illustrates the usefulness of our system, in terms of conservation
of motifs. We illustrate here the functional features of kinases,
also present in the BaliBase as well as used for illustration by
Notredame~\cite{tcoffee}. Each sequence is identified by its
SwissProt/UniProt identifier. Some of the identifications have
changed, and are illustrated as is viewed in SwissProt at the time
of this publication. In the example, there are 3 motifs
identified. These motifs are the core blocks identified by
BaliBase, and are conserved by Sample-Align-D, marked as red,
orange and blue. The motifs are in greater order of difficulty,
with red as the least difficult. The blue labeled motif is the
most difficult to conserve in the example set, because of the long
indel in KIN3-Yeast. As can be seen, Sample-Align-D is able to
conserve this most difficult motif, for decomposition factor of 4
as done for benchmarks.


\subsection{Performance in Terms of HPC Parameters}

In this section we analyze performance in terms of execution time,
scalability, and memory. The objective of the evaluation is to
determine the advantages of the proposed domain decomposition
based technique in terms of speedup and reduction in memory
requirements. In this section we also compare Sample-Align-D with
the known parallel approaches reported in the literature.

\subsubsection{Execution Time and Memory}

%
%
For the sake of coherence in our presentation we generated
sequences with same parameters that were used for the quality
assessment. We report results for up to 20000 sequences. To the
best of authors' knowledge, there are no published reports of
aligning this large number of sequences in the literature.

\begin{figure}[htb]
\begin{center}
\includegraphics[scale=0.5,angle=0]{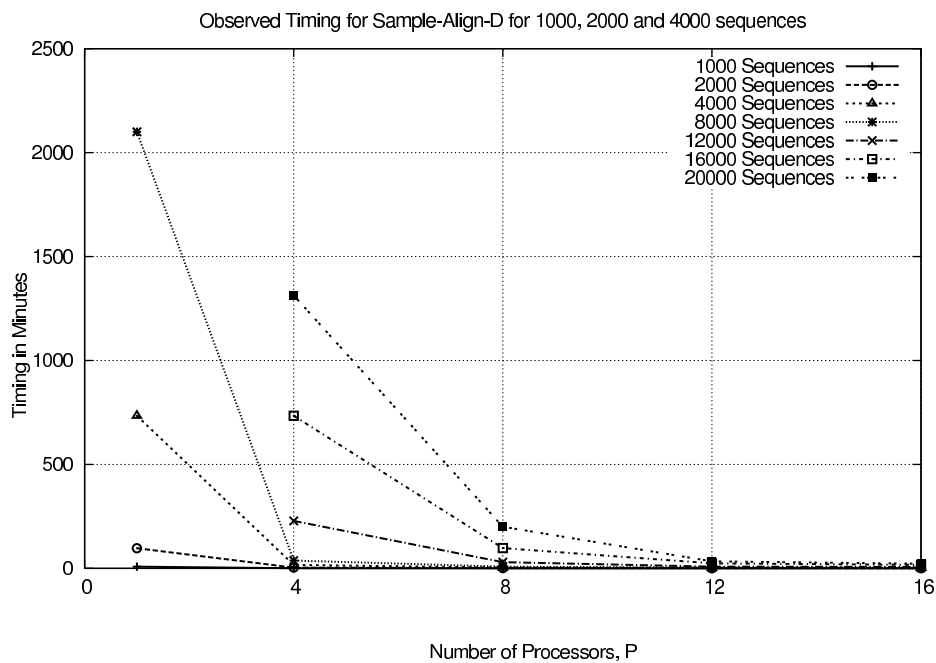}
\caption{\small \label{fig-timing} Scalability of the execution
time w.r.t. the number of processors.}
\end{center}

\end{figure}

As shown in Fig.~\ref{fig-timing}, in the case of Sample-Align-D
the execution time decreases sharply with the increase in the
number of processors. We are able to align 8000 sequences in just
3.9 minutes, compared to 2100 minutes on a sequential Muscle
System. The timing for 12000, 16000 and 20000 sequences are also
shown for Sample-Align-D. The timing for one node with Muscle is
not shown because Muscle Systems was not able to handle this large
number of sequences and the resources requirement in terms of
memory and time, grew exponentially for these sets of sequences.

\begin{figure}[htb]
\begin{center}
\includegraphics[scale=0.5,angle=0]{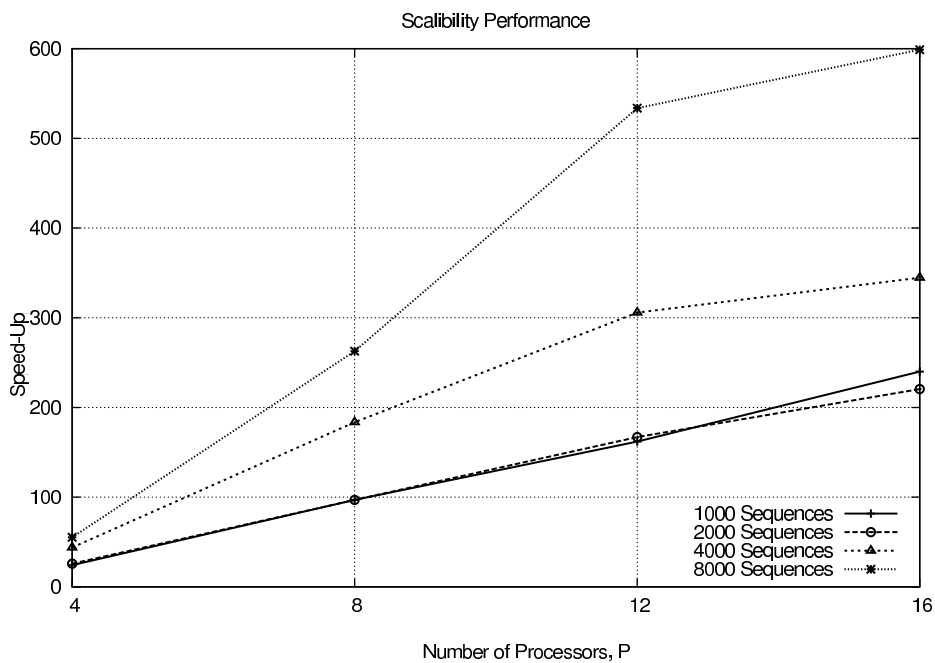}
\caption{\small \label{fig-speedup} Super-linear speed-ups for
Sample-Align-D with increasing number of processors.}
\end{center}
\end{figure}

As shown in Fig.~\ref{fig-speedup}, Sample-Align-D Algorithm
exhibits super linear speed-ups (of the order of 600) on a 16
processor system. This is primarily because the computation
complexity decreases by $O(p^4)$ with the increase in the number
of processor, as suggested in our algorithmic analysis section.

\begin{figure}[htb]
\begin{center}
\includegraphics[scale=0.4,angle=0]{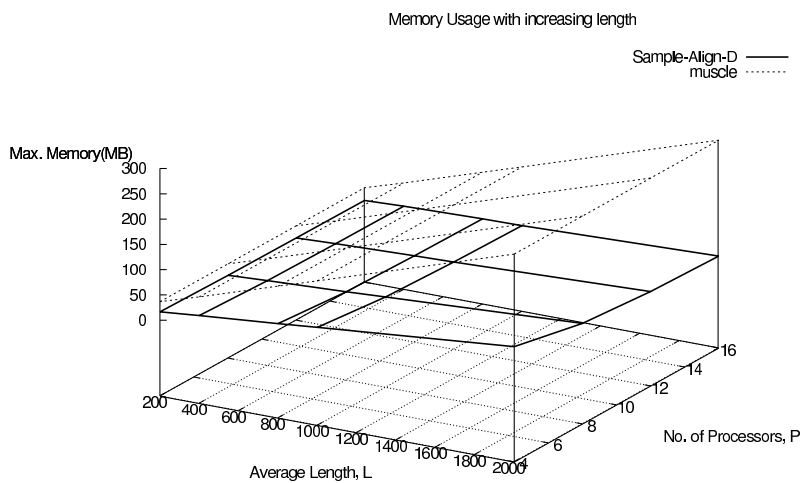}
\caption{\small \label{fig-result2} Memory usage, with varying
average length of the sequences.}
\end{center}
\end{figure}

\begin{figure}[htb]
\begin{center}
\includegraphics[scale=0.4,angle=0]{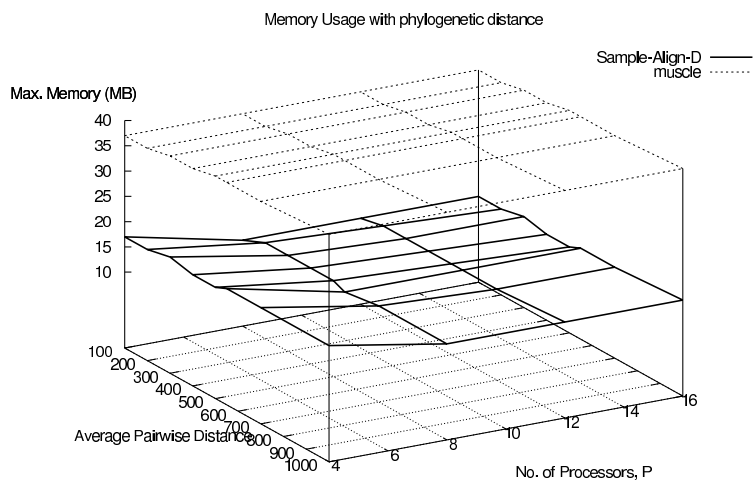}
\caption{\small \label{fig-result6} Memory usage, with varying
average pairwise distance of sequences.}
\end{center}
\end{figure}

\begin{figure}[htb]
\begin{center}
\includegraphics[scale=0.4,angle=0]{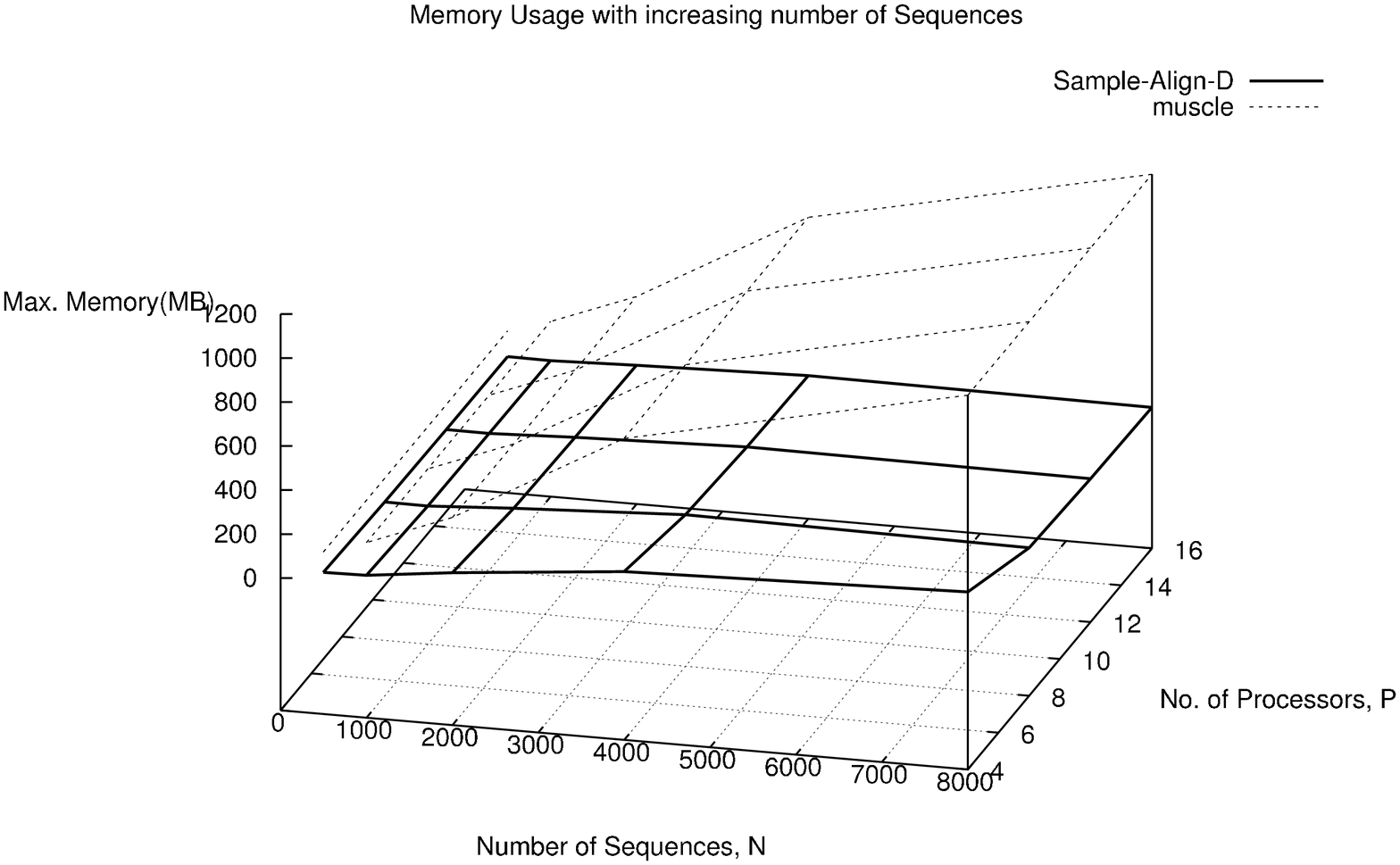}
\caption{\small \label{fig-result9} Memory usage, with varying
number of sequences.}
\end{center}
\end{figure}

The low memory requirements, as predicted by our analysis in
Section~\ref{sec-analysis}, are also evident in our experimental
results. The memory requirements while increasing the length of
sequences, phylogenetic distance, and number of sequences, are
shown in Fig.~\ref{fig-result2} Fig~\ref{fig-result6} and
Fig.~\ref{fig-result9} respectively. The most interesting figure
is the one that depicts the memory requirements with increasing
number of sequences. As can be seen in Fig.~\ref{fig-result9},
with the increase in the number of sequences, the memory
requirements for Muscle System is increasing exponentially with
1200 MB required for 8000 sequences. However, the maximum memory
required for Sample-Align-D even for 8000 sequences is not greater
than 100 MB.

\subsubsection{Comparison with Existing Parallel MSA Systems}

There have been significant efforts towards parallelizing MSA
techniques, as discussed in Section 3. We have selected Parallel
Clustalw and Parallel T-Coffee, two of the most widely used
parallel MSA systems, to compare the performance of the proposed
Sample-Align-D Algorithm. The limitation of this aspect of
evaluation was also the selection of the common sequence set. This
is due to the fact that most of the existing parallel systems are
unable to handle large number of sequences because of one or two
explicit sequential stages in these solutions.

\begin{figure}[htb]
\begin{center}
\includegraphics[scale=0.5,angle=0]{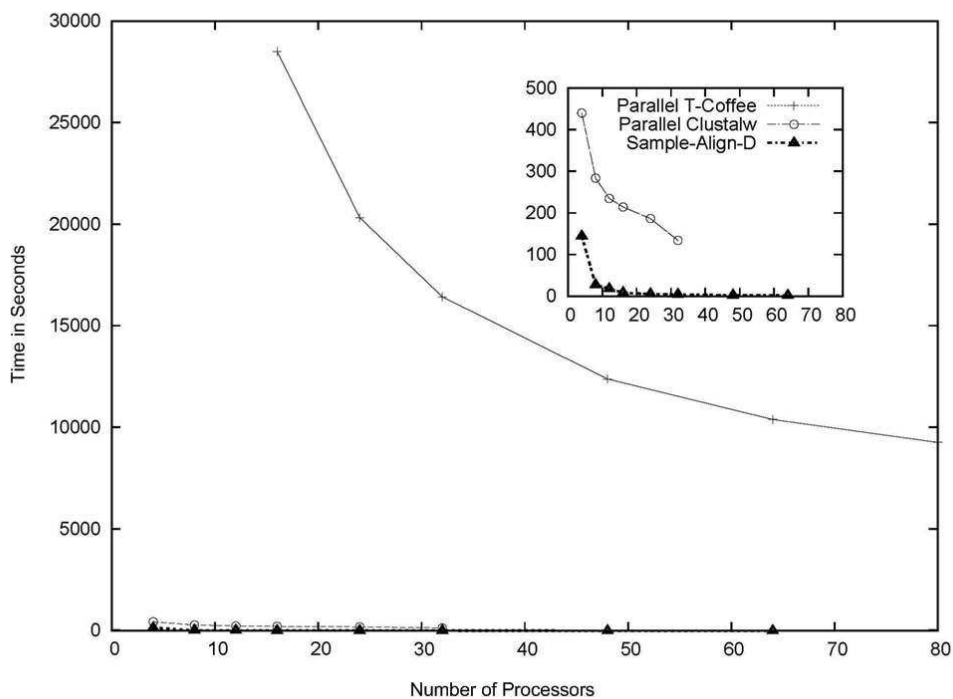}
\caption{\small \label{fig-comparsion} Comparison of execution
times of parallel T-Coffee, parallel Clustalw and Sample-Align-D}
\end{center}
\end{figure}

We chose the data sets that Zola et al.~\cite{tcoffeep} used for
the evaluation of Parallel T-Coffee, named PF00500, consisting of
1048 sequences with maximal length of 523 characters. The
execution times of different parallel algorithms for the data sets
are plotted in Fig.~\ref{fig-comparsion}. The execution time of
Parallel T-Coffee is significantly higher than that of the
Sample-Align-D algorithm. For example, on a 16 processor system,
it took around 9.1 hours for Parallel T-Coffee, 4.3 minutes for
Parallel Clustalw, and only 8.1 seconds for Sample-Align-D. Our
experiments show that the performance of Parallel Clustalw
degrades significantly compared to Sample-Align-D as the number of
sequences in the set increases.

\section{Conclusion and Discussions}

We have described a domain decomposition (data parallel) strategy
for multiple sequence alignment of biological sequences. To our
knowledge, this is the first attempt to investigate domain
decomposition for the multiple sequences alignment problem. This
domain decomposition allowed us to devise a highly scalable
multiple alignment system. A detailed algorithmic technique based
on a novel decomposition strategy was described and rigorous time
and space complexity analyses were presented. The proposed
strategy decreased the time complexity of any MSA heuristic
$O(N)^x$ by a factor of $O(1/p)^x$. Consequently, super-linear
speed-ups were achieved and tremendous decrease in memory
requirements were observed, as predicted by the complexity
analysis. A rigorous quality analysis of the decomposition
technique was also introduced, and the effect of decomposition on
the quality of the alignment was investigated. The quality
analysis allowed us to determine the quality of the alignment
relative to that of the underlying sequential MSA system.

A number of research problems remain open and the techniques
introduced in this manuscript suggest new directions of research
that can be pursued. The open research problems in computational
biology and parallel processing that arise from the research
presented in this manuscript are as follows:

\begin{enumerate}

\item We have presented the decomposition strategy as a parallel
computing solution. However, the super-linear speed-ups on
multiple processors suggest that the use of the sampling based
decomposition strategy on single processor systems would also be
able to deliver significant time and space advantages as evident
in~\cite{SaeedKhokhar2}. The decomposition strategy in any of the
existing multiple alignment systems with some form of iterative
quality 'patch-up' strategy would be useful. Also, multi-core
processors can take enormous advantage of the decomposition
strategy for improved efficiency.

\item  In our partitioning strategy, the load balancing scheme is
based on the k-mer rank and the number of sequences. It would be
interesting to develop a more elaborated load balancing scheme
considering additional factors such as the length of the sequences
and the type of sequences being considered.

\item The domain decomposition based strategy has been
investigated for distance based multiple sequences alignment
methods such as Muscle and Clustalw. It would be interesting to
investigate the same or similar partitioning strategy for other
type of consistency and profile based methods such as T-Coffee,
ProbCons, Mafft, DbClustal, MUMMALS etc.

\item We have considered a subset of alignment parameters, for
example, PSP scores, 200 PAM matrix and the 240 PAM VTML matrix.
It would be insightful to consider different mutation matrix and
other parameters, and investigate the effects of decomposition on
quality.

\item Phylogenetic trees are the crux of research on evolutionary
biology. However, building phylogenetic trees for large number of
species is considerably compute intensive. It would be useful to
apply decomposition to build distance based phylogenetic trees for
multiple genomes.

\item As the sizes of the biological sequence archives and
structural data are increasing at an exponential rate, the pattern
and motifs searching is getting increasingly more time consuming.
The application of decomposition strategy could allow to search
these keys motifs in such databases. The strategy can also find
its applications in target and lead identification in drug
discovery.

\item The strategy can be used to obtain computational analysis
without having the need to actually 'import' the sequences locally
e.g. multiple sequences alignment can be performed on distant
databases, without transferring the entire set of sequences, which
some times might be desirable due to proprietary data issues etc.

\end{enumerate}

\clearpage

\begin{center}
    {\bf APPENDIX}
\end{center}
\begin{algorithm}
\caption{Sample-Align-D (sequences $N$)}\label{algo2}
\begin{algorithmic}
\scriptsize{ \REQUIRE $p$ processors for computation and $N$
sequences of amino acids $S_1, S_2, \cdots, S_N$:} \ENSURE
\scriptsize{Gaps are inserted in each sequence  such that:}
\begin{itemize}
\item \scriptsize{All sequences have the same length and the Score
of the global map is maximized according to the chosen scoring
function}
\end{itemize}

\begin{enumerate}

\item \footnotesize{Assume $N/p$ sequences on each of the $p$
processors}

\item Locally compute k-mer rank of all the sequences in each
processor

\item Sort the sequences locally in each processor based on k-mer
rank

\item Choose a sample set of $k$ sequences in each processor,
where $k\ll N/p$

\item Send the $k$ samples from each processor to all the
processors.

\item Compute the k-mer rank of each sequence against the $k
\times p$ samples.

\item Sort the sequences locally in each processor based on the
new k-mer rank.

\item Using regular sampling, choose $p-1$ sequences from each
processor and send only their ranks to a root processor.

\item Sort all the $p \times(p-1)$ ranks at the root processors
and divide the range of ranks into $p$ buckets.

\item Send the bucket boundaries to all the processors.

\item Redistributed sequences among processors such that sequences
with k-mer rank in the range of bucket $i$ are accumulated at
processor $i$, where $0> i < p+1$.

\item Align sequences in each processor using any sequential
multiple alignment system

\item Broadcast the Local Ancestor to the root processor

\item Determine Global Ancestor GA at the root processor by
aligning local ancestors received from all the processors

\item Broadcast GA to all the processors

\item Realign each of the sequences in $p$ processors based on
ancestor GA using profile-profile alignment i.e. Each of the
profiles of aligned sequences are tweaked using the ancestor
profile, with constraints.

\item Glue all the aligned sequences at the root processor.
\end{enumerate}

\end{algorithmic}
\end{algorithm}


\clearpage
\bibliography{mybib}
\bibliographystyle{ieeetr}
\end{document}